# Beyond Tool Adoption: A Practical Five-Stage Developmental Continuum for AI Literacy in Higher Education

J. Paul Liu[1,2,3] and Rachel Levy[4,5]

1. Dept of Marine, Earth, and Atmospheric Sciences; 2. AI Hub for Science; 3. Center of Geospatial Analytics; 4. Data Science and AI Academy; 5. Department of Mathematics

North Carolina State University

Corresponding author: jpliu@ncsu.edu; rlevy@ncsu.edu

## Abstract

Artificial intelligence (AI) literacy is increasingly recognized as a foundational competency for all university graduates. Yet students' engagement with AI tools often clusters at two extremes: avoidance driven by fear, mistrust, ethical concern, or lack of access, and uncritical reliance that produces fluent output while masking misunderstanding. Existing AI literacy frameworks provide valuable competency definitions, but most offer limited guidance for diagnosing where learners begin and how they progress toward responsible, critical engagement. This paper proposes **a five-stage AI Literacy Continuum**–0) Not Yet Engaged, 1) Uncritical Use, 2) Informed Use, 3) Critical Evaluation, and 4) Improvement–that describes developmental orientations toward AI use in higher education. The continuum complements dimensional frameworks by **providing educators with a practical diagnostic and instructional pathway** aligned with international frameworks, including UNESCO and OECD. We present a design-based implementation case from North Carolina State University, where credit-bearing courses and intensive hands-on workshops engaged more than 330 participants between Fall 2024 and Spring 2026. Because the implementation did not use a validated pre/post instrument or comparison group, we frame the findings as observational and practice-based: participants exhibited behaviors consistent with movement from non-engagement or uncritical use toward informed engagement, while sustained and discipline-embedded experiences produced stronger evidence of critical evaluation and improvement-oriented practice. We discuss curricular pathways, opportunity considerations, assessment strategies, and argue **that AI literacy should be understood not as tool adoption alone but as a developmental capacity to understand, evaluate, and responsibly apply AI systems in disciplinary and societal contexts.**

**AI use disclosure:** The authors used generative AI tools (ChatGPT/Claude) for language editing, formatting of tables, and prose refinement. All conceptual content, data, analysis, references, and conclusions are solely the work of the authors. No AI tools were used for idea generation or the synthesis of findings.

## 1. Introduction

Artificial intelligence has become deeply embedded in knowledge production, professional practice, and everyday decision-making. From scientific discovery and engineering design to writing, coding, legal analysis, and data interpretation, AI systems increasingly mediate how students learn and how they demonstrate understanding (Kasneci, et al., 2023; Liu, 2026; Yan et al., 2024). The rapid proliferation of generative AI tools, particularly large language models such as ChatGPT, Claude, and Gemini, has accelerated this integration, making AI interaction a near-daily reality for university students across disciplines. As a result, universities and national education systems are converging on a shared objective: all graduates should possess baseline AI literacy, comparable in importance to digital literacy or information literacy (Long & Magerko, 2020; Ng et al., 2021) (Supplementary Table 1).

International organizations have formalized this aspiration. UNESCO's AI Competency Framework for Students identifies twelve competencies organized across four dimensions: human-centered mindset, ethics of AI, AI techniques and applications, and AI system design across three progression levels: Understand, Apply, and Create (UNESCO, 2024). The OECD and European Commission have jointly developed an AI Literacy Framework for primary and secondary education, with implications extending into higher education (OECD, 2025). National initiatives, institutional strategic plans, and accreditation standards increasingly reference AI literacy as an expected graduate outcome.

In practice, however, this policy aspiration collides with heterogeneous and often problematic patterns of student engagement. Drawing on multi-year instructional practice across AI and generative AI courses and intensive training workshops with several hundred university students, a consistent pattern emerges: student engagement with AI tools tends to cluster at **two extremes**. At one pole students avoid AI tools entirely, driven by fear of academic integrity violations, mistrust of AI-generated content, ethical concerns about resource usage, data privacy and labor displacement, or simply lack of prior exposure. At the other pole, students adopt generative AI rapidly and uncritically, treating AI-generated outputs as authoritative and submitting work that reflects fluent AI prose rather than genuine understanding. This second pattern is particularly concerning because it produces what has been described as "performance without learning"—student work that appears competent on the surface but masks fundamental gaps in disciplinary knowledge and critical reasoning (Pinski & Benlian, 2024).

This polarization reveals a structural gap in how AI literacy is conceptualized and taught. Most existing frameworks define AI literacy through competency dimensions—cataloging what students should know and be able to do across domains such as technical understanding, ethical reasoning, and practical application. **Dimensional frameworks** such as the AI Literacy Heptagon (Hackl, et al., 2025), ED-AI Lit (Allen & Kendeou, 2024), and the Digital Education Council framework (DEC, 2025) provide valuable taxonomies of AI literacy competencies. However, they offer limited guidance on how students develop along these dimensions over time. Educators therefore face practical challenges: How do we diagnose where individual students are in their AI literacy development? How do we design instruction that moves learners forward from teir current state? How do we assess

AI literacy in ways that capture genuine understanding rather than surface performance? How can we help students develop awareness of their progress?

Developmental models such as those that describe sequential progression through identifiable stages of competence, address these questions more directly. UNESCO's three-level progression (Understand, Apply, Create) represents one such approach, as does the competency-based ladder pathway proposed for vocational students (Hong, 2025). However, these models either operate at high levels of abstraction or target populations outside higher education. A developmental framework calibrated to the specific patterns of engagement observed among university students remains needed.

The specific gap addressed here is not simply that existing frameworks are insufficiently developmental. It is that they rarely name two highly visible pre-literacy patterns in current classrooms: principled or uninformed avoidance of AI, and fluent but uncritical AI use. These two patterns create different instructional needs. Avoidance requires access, demystification, and ethical nuance; uncritical use requires verification habits, metacognitive awareness, and disciplinary judgment. A developmental continuum helps educators distinguish these conditions rather than treating all AI exposure as progress.

**This paper addresses that gap by proposing a five-stage AI Literacy Continuum grounded in observed patterns of student engagement in higher education**. The continuum provides a shared language for educators, administrators, and students, enabling scalable curriculum design and assessment while remaining aligned with international competency frameworks. We present a conceptual framework supported by design-based observational evidence from institutional implementation. We explicitly identify formal empirical validation as a priority for future research.

The remainder of this paper is organized as follows. Section 2 reviews related work on AI literacy definitions, frameworks, and measurement. Section 3 presents the five-stage continuum with illustrative examples. Section 4 maps the continuum to existing frameworks. Section 5 discusses educational design implications. Section 6 addresses assessment strategies. Section 7 presents a case study of AI literacy implementation at North Carolina State University, providing observational evidence that the proposed continuum stages are practically achievable. Section 8 discusses contributions, limitations, and future directions. Section 9 concludes.

# 2. Related Work

## 2.1 Defining AI Literacy

AI literacy has been defined in multiple ways across disciplines and has evolved substantially over the past decade. In the 2010s, the focus largely centered in the tech/STEM community on machine learning to solve image processing, classification/clustering, anomaly detection and recommendation systems. The Netflix Prize galvanized this work and brought interdisciplinary teams together to work with massive training and test data sets with new algorithmic approaches.

Early literacy concepts emphasized understanding basic AI definitions and practices—what AI is, how it differs from conventional software, and recognizing AI applications in daily life (Long & Magerko, 2020). Concurrently, the AI4K12 Initiative identified five "Big Ideas" in AI—Perception, Representation and Reasoning, Learning, Natural Interaction, and Societal Impact—as an organizing framework for K–12 AI education (Touretzky et al., 2019), which has since been widely adopted internationally. These foundational ideas were valuable in establishing AI literacy as a distinct construct but focused primarily on awareness and recognition.

More recent definitions reflect the increasing complexity and ubiquity of AI systems. An exploratory review published in *Computers and Education: Artificial Intelligence*, proposed four interrelated aspects of AI literacy: know and understand AI, use and apply AI, evaluate and create AI, and understand ethical issues (Ng et al., 2021). This framework drew on analogies with digital literacy and information literacy, establishing AI literacy as a multi-dimensional construct rather than a single skill. Allen and Kendeou (2024) extended this work through the ED-AI Lit framework, which grounds AI literacy in learning sciences and cognitive psychology. Their six-component model—Knowledge, Evaluation, Collaboration, Contextualization, Autonomy, and Ethics—emphasizes that AI literacy should support meaningful learning rather than mere automation of tasks.

A systematic review of the evolving landscape of AI literacy research spanning 2014 to 2024 documents the rapid expansion of the field, noting increasing attention to critical evaluation, ethical reasoning, and the implications of generative AI for literacy definitions (Yang et al., 2025). Earlier scoping reviews in higher education (Laupichler et al., 2022) and K–12 settings (Casal-Otero et al., 2023) similarly identified growing but fragmented attention to AI literacy, with limited consensus on operational definitions. This evolution reflects a broader shift away from viewing AI literacy as understanding technology and toward viewing it as the capacity to engage with a trustworthy AI system as an informed, critical, and responsible agent.

## 2.2 Dimensional Frameworks

The predominant organizational approach in current AI literacy scholarship is dimensional: frameworks that identify distinct competency areas or domains that collectively constitute AI literacy. These frameworks answer the question, "What should an AI-literate person know and be able to do?"

The AI Literacy Heptagon (Hackl, et al., 2025) exemplifies this approach, identifying seven equally weighted dimensions: Technical, Applicational, Critical Thinking, Ethical, Social, Integrational, and Legal. The model's distinctive contribution is the inclusion of integrational and legal dimensions, which are often underrepresented in earlier frameworks. The heptagonal visualization emphasizes that no single dimension should dominate, addressing concerns that AI literacy may become overly technical at the expense of ethical and social understanding.

The Digital Education Council AI Literacy Framework (DEC, 2025) proposes five dimensions: Understanding AI and Data, Critical Thinking and Judgment, Ethical and Responsible AI Use, Human-Centricity and Emotional Intelligence, and Domain Expertise. Similarly, the EDUCAUSE framework (2024) organizes AI literacy for higher education around four key areas: Technical Understanding, Evaluative Skills, Practical Application, and Ethical Considerations.

These dimensional frameworks provide essential conceptual infrastructure. They establish the scope and boundaries of AI literacy, support the development of learning outcomes, and facilitate cross-institutional comparison. However, they share a common limitation: they specify the target state of AI literacy without describing how learners reach that state. For educators seeking to design curricula that meet students where they are and guide them forward, dimensional frameworks alone provide insufficient guidance.

## 2.3 Developmental and Progression-Based Approaches

A growing body of work recognizes that AI literacy develops progressively and that stage-based or leveled models can complement dimensional approaches.

UNESCO's AI Competency Framework for Students (2024) represents the most prominent progression-based framework, organizing its twelve competencies across three developmental levels—Understand, Apply, and Create—that describe increasing sophistication of engagement. This structure implicitly acknowledges that AI literacy is not binary (literate or illiterate) but exists on a continuum.

The OECD and European Commission AI Literacy Framework (2025) distinguish competency expectations across primary and secondary education levels, tying developmental distinction to educational stage. Chee et al. (2025), in a competency framework published in the *British Journal of Educational Technology*, explicitly vary competency expectations across learner groups—K-12 students, higher education students, and workforce professionals—with implied learning pathways between groups. In the domain of generative AI specifically, frameworks have emerged that organize competencies along progression hierarchies following Bloom's cognitive taxonomy (Liu et al., 2025). The competency-based ladder pathway for vocational students (Hong., 2025) proposes a three-tier progression: foundational cognitive, skills application, and comprehensive innovation.

In higher education, aligning with a well-known framework such as Bloom's also helps academic institutions recognize these new literacy components in the context of a common framework that is likely used, for example, by curriculum committees to develop and evaluate learning objectives and other syllabus components.

These developmental approaches align with established traditions in educational theory. Bloom's taxonomy (Bloom et al., 1956; Anderson & Krathwohl, 2001), the Dreyfus model of skill acquisition (Dreyfus & Dreyfus, 1986), and Perry's scheme of intellectual development (Perry, 1970) all demonstrate that learning progressions can be usefully described as movement through identifiable stages of increasing sophistication. What remains lacking is a developmental model calibrated specifically to the patterns of AI engagement observed in higher education—one that accounts for both the avoidance and the uncritical adoption that characterize current student behavior.

## 2.4 AI Literacy in Higher Education

Scoping and systematic reviews indicate rapid growth in AI literacy initiatives across higher education, particularly following the widespread availability of generative AI tools in late 2022 and 2023. However, these reviews also highlight substantial challenges. Pinski and Benlian (2024) identify uneven implementation across disciplines, limited assessment practices, and a persistent tendency to focus on tool exposure rather than evaluative competence. Ng et al. (2024) observe that many initiatives lack theoretical grounding and treat AI literacy as a byproduct of tool use rather than a deliberate educational outcome.

Concerns have emerged that generative AI may exacerbate rather than resolve language and concept literacy gaps. When students can produce fluent, well-structured outputs by prompting an AI system, traditional assessments may fail to distinguish genuine understanding from AI-assisted performance. This dynamic creates an urgent need for frameworks that help educators differentiate between students who use AI skillfully and those who merely use AI to reduce intellectual burden.

Note that this may be an important distinction between appropriate use of AI for the purposes of learning by a novice versus as an efficiency tool by an expert. AI technologies have roles to play in both contexts, but we are still discovering the most effective methodologies for AI use in these contexts that have different goals. Research on productive struggle and productive failure (Kapur, 2008; Kapur & Bielaczyc, 2012) and on desirable difficulties in learning (Bjork, 1994) may lend insight about the need to not reduce intellectual challenge too much during a learning phase in order to build what mathematician Weiqing Gu calls intellectual or "mathematical muscle," analogous to making sure muscles build mass through challenge, whether it be gravity or extra weight at the gym. When AI reduces the cognitive effort required to produce outputs, learners may experience what Fan et al. (2025) term "metacognitive laziness"—a pattern in which AI-assisted task performance improves but metacognitive engagement (monitoring, evaluation, orientation) declines, ultimately hindering knowledge gain and transfer.

## 2.5 Measuring AI Literacy

Several validated measurement instruments now exist for assessing AI literacy across populations. A systematic quality assessment of 16 validated scales using COSMIN criteria found generally good structural validity and internal consistency, though few scales had been tested for cross-cultural validity or measurement error (Lintner, 2024). The Scale for the Assessment of Non-Experts' AI Literacy (SNAIL) provides a 31-item measure organized around technical understanding, critical appraisal, and practical application (Laupichler et al., 2023). The Meta-AI Literacy Scale (MAILS) captures facets including using and applying AI, understanding AI, detecting AI, AI ethics, and creating AI (Carolus et al., 2023). SAIL4ALL offers 56 items across four themes and is notable as a performance-based instrument suitable for general populations (Soto-Sanfiel et al., 2025).

More recently, instruments specific to generative AI have been developed, including ChatGPT literacy scales and the Generative AI Literacy Assessment Test (GLAT). Despite their psychometric value, these instruments are rarely integrated into curricular frameworks that would guide their use in instructional design and assessment planning. A developmental framework can help bridge this gap by clarifying which instruments are appropriate for assessing which stages of AI literacy.

## 3. The AI Literacy Continuum

To bridge the gap between abstract competency definitions and classroom practice, this section proposes a five-stage AI Literacy Continuum (**Figure 1**). The continuum describes typical patterns of learner engagement with AI systems in higher education, moving from complete avoidance through increasingly sophisticated and responsible use.

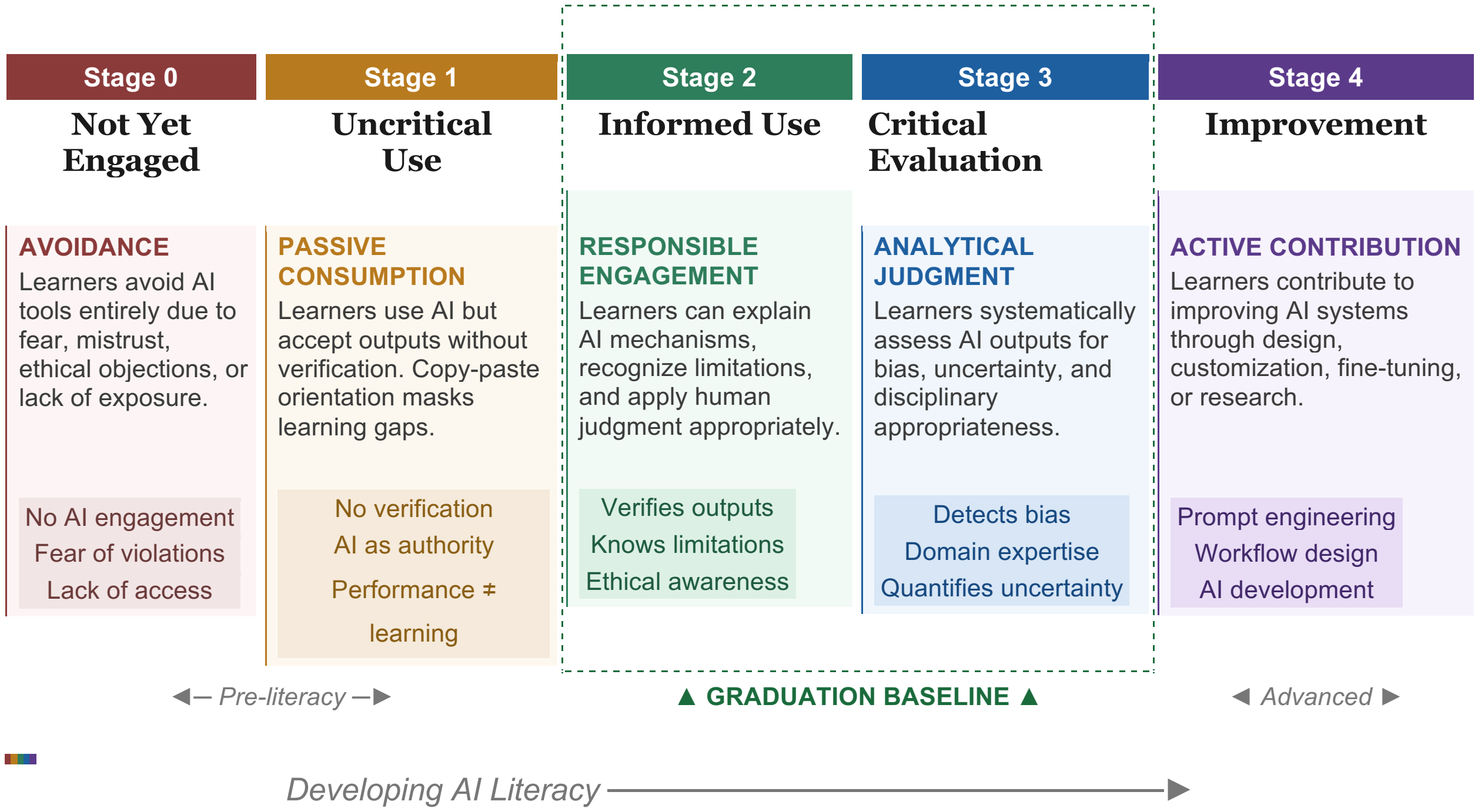

**Figure 1.** The AI Literacy Continuum (Stages 0–4). The dashed border indicates the recommended graduation baseline for higher education (Stages 2–3). Progression is generally sequential but not strictly linear; learners may exhibit behaviors across multiple stages depending on context.

Several design principles guide the continuum. First, the stages describe predominant orientations toward AI rather than rigid categories; individual students may exhibit behaviors associated with multiple stages depending on context, task, or disciplinary domain. The continuum therefore describes orientations toward particular forms of AI engagement rather than stable personal characteristics. For example, a student might operate at Stage 3 (Critical Evaluation) when analyzing historical texts but engage at Stage 1 (Uncritical Use) when writing Python code, highlighting that AI literacy is often domain-dependent. Second, progression is generally sequential but not strictly linear—students may regress under unfamiliar conditions or advance rapidly with appropriate instruction. Third, the continuum is descriptive rather than normative at the individual level: Stage 0 is not inherently inferior to Stage 4 for every purpose, though institutional goals should generally target Stages 2–3 as graduation baselines.

**Table 1. Diagnostic indicators and assessment examples for the AI Literacy Continuum.** The indicators below operationalize the continuum by distinguishing observable behaviors from possible assessment evidence. They are practical heuristics rather than psychometrically validated cut scores.

| Stage | Dominant Orientation | Observable Indicators | Possible Assessment Evidence |
|---|---|---|---|
| **Stage 0** – Not Yet Engaged | No meaningful engagement with AI | • Non-use of AI tools<br>• Limited or no exposure<br>• Fear of academic integrity violations<br>• Lack of access<br>• Ethical refusal to use AI | • Entry surveys<br>• Reflective statements<br>• Discussions of ethical concerns<br>• Documentation of access barriers |
| **Stage 1** – Uncritical Use | Acceptance of AI outputs with minimal scrutiny | • Treats AI outputs as authoritative<br>• Copies or lightly edits outputs<br>• Fails to verify information<br>• Omits disclosure of AI use | • Exercises identifying hallucinated citations<br>• Detection of unsupported claims<br>• Verification activities |
| **Stage 2** – Informed Use | Appropriate and responsible use of AI | • Selects suitable AI applications<br>• Verifies facts and sources<br>• Recognizes hallucinations and bias<br>• Documents AI use | • AI-use process logs<br>• Annotated outputs<br>• Rationales for use decisions |
| **Stage 3** – Critical Evaluation | Analytical evaluation of AI-generated content | • Evaluates assumptions and uncertainty<br>• Assesses bias and disciplinary fit<br>• Considers consequences of error<br>• Critiques strengths and limitations | • Discipline-specific critiques<br>• Comparative analyses<br>• Reflective evaluations |
| **Stage 4** – Improvement | Creation and enhancement of AI systems and workflows | • Designs improved prompts and workflows<br>• Develops evaluation protocols<br>• Curates datasets<br>• Builds/customizes AI systems | • Capstone projects<br>• Prompt workflow designs<br>• Retrieval pipelines<br>• Evaluation rubrics<br>• Model-improvement reports |

## 3.1 Stage 0: Not Yet Engaged with AI

At Stage 0, learners have not yet explored AI tools or are unaware of them. This lack of exposure and experience may stem from multiple sources: fear of academic integrity violations ("using AI is cheating"), mistrust of AI-generated content, principled ethical objections regarding data privacy or labor displacement, institutional prohibitions, lack of access to technology or simply a lack of prior exposure to AI systems.

Consider a graduate student in the sciences or humanities who refuses to use any AI tools for research or writing, believing that AI-generated text is fundamentally incompatible with scholarly integrity. While this position may reflect thoughtful ethical reasoning, it also limits the student's ability to engage with AI-augmented research workflows that are becoming standard in their field. Similarly, other students may arrive at university with minimal exposure to AI tools, not through active refusal but through lack of access.

It is important to distinguish principled refusal from uninformed avoidance. Both result in non-engagement with AI, but they require different instructional responses. Principled refusers may benefit from nuanced discussions of AI ethics and appropriate use, while students lacking exposure need structured introduction and access. In both cases, remaining at Stage 0 may increasingly limit graduates' preparedness for AI-integrated professional and civic life.

We recognize that some individuals and groups consider Stage 0 an ideal state because of ethical or environmental concerns. Teaching and learning environments must be clear about what options, if any, are available for learners who wish to minimize their use of AI.

Importantly, Stage 0 should not be interpreted as intellectual deficiency. Some learners may arrive at this stage through informed ethical or environmental concerns rather than lack of knowledge. The continuum describes engagement with AI systems rather than overall technological sophistication.

## 3.2 Stage 1: Uncritical Use

At Stage 1, learners use AI tools but accept outputs without verification or critical examination. This stage is characterized by what might be termed a "copy-paste" orientation: the student treats AI-generated content as authoritative, substituting AI outputs for their own analysis, reasoning, or synthesis.

A common manifestation occurs when a science student uses a large language model to generate a laboratory report, submitting the AI-produced text without checking whether the reported data, citations, or interpretations are accurate. The resulting work may appear polished and competent—generative AI systems produce fluent, well-structured prose—but the apparent quality masks the student's lack of engagement with the underlying material. This pattern represents "performance without learning": the student has completed the task but has not developed the disciplinary understanding the task was designed to build. Empirical evidence supports this concern: Fan et al. (2025) demonstrated in a randomized experiment that ChatGPT use improved short-term essay scores but triggered "metacognitive laziness," with students showing reduced monitoring, evaluation, and orientation behaviors and no significant gains in knowledge transfer.

Stage 1 is particularly difficult to address because it can be difficult to detect through conventional assessment. A well-crafted AI-generated essay may receive a high grade under traditional rubrics that evaluate structure, coherence, and surface-level argumentation. The student's confidence may also be inflated, creating a metacognitive gap between perceived and actual understanding. AI detection algorithms are documented to fail, to including false positives. Flags on student work can lead to time consuming and taxing engagements with academic integrity boards.

Instructionally, Stage 1 represents a more urgent concern than Stage 0. Students who have not yet engaged with AI can be gradually introduced to its benefits; students who use AI uncritically may actively resist moving toward more effortful critical engagement, since their current approach already produces satisfactory outcomes.

## 3.3 Stage 2: Informed Use

At Stage 2, learners understand basic mechanisms of AI systems, recognize appropriate use cases and limitations, and use AI responsibly with awareness of when human judgment is required.

An informed user might employ a large language model to generate an initial literature review draft but then systematically verify each cited reference, cross-checking against library databases and correcting hallucinated citations. They understand that language models generate probabilistic text rather than retrieving verified facts, and they adjust their workflow accordingly. They can articulate when AI assistance is appropriate (brainstorming, initial drafting, code debugging) and when it is not (making clinical decisions, evaluating evidence quality, ethical reasoning). Appropriate usage will likely vary by discipline and even by type of activity.

Stage 2 represents the minimum graduation baseline for AI literacy in higher education. At this stage, ethical awareness becomes explicit: students recognize concerns about bias in training data, intellectual property considerations, and the importance of transparency about AI use. The transition from Stage 1 to Stage 2 is the most educationally significant, as it marks the shift from passive consumption to active, informed engagement.

## 3.4 Stage 3: Critical Evaluation

At Stage 3, learners systematically verify AI outputs, recognize bias and uncertainty, and contextualize AI-generated results within disciplinary knowledge and broader societal implications. While users at stage 2 gain awareness of these issues, they may not yet be able effectively to address and mitigate them.

Critical evaluators go beyond checking whether AI outputs are factually correct. They assess the reasoning process, identify potential sources of bias in training data and model design, quantify uncertainty, and evaluate whether AI-generated solutions are appropriate for the specific disciplinary context. For example, an environmental science student using AI-assisted data analysis might not only verify the statistical outputs but also evaluate whether the model's assumptions are appropriate for the specific ecosystem being studied, consider what data might be missing or underrepresented in the training set, and articulate the limitations of the AI analysis in the broader context of environmental policy.

Stage 3 engagement generally requires substantial disciplinary knowledge. Critical evaluation of AI outputs is domain-dependent: one cannot assess whether an AI-generated medical diagnosis is reasonable without medical knowledge, nor evaluate an AI-generated legal analysis without legal expertise. Distinguishing Stage 2 from Stage 3 is particularly important. Learners at Stage 2 can recognize AI limitations, verify outputs, and select appropriate use cases. Learners at Stage 3 go further by evaluating assumptions, uncertainty, disciplinary appropriateness, and consequences of error. For example, a Stage 2 learner may verify whether an AI-generated citation exists, whereas a Stage 3 learner may critique whether the cited evidence is representative, sufficient, and appropriate for the disciplinary context (Table 2).

**Table 2. Illustrative Distinction Between Stages 2 and 3**

| Task | Stage 2 | Stage 3 |
|---|---|---|
| AI-generated essay | Verifies sources | Evaluates evidentiary quality and bias |
| AI-generated code | Tests functionality | Evaluates assumptions, efficiency, reliability |
| Literature review | Checks citations | Critiques coverage and omissions |
| Data analysis | Verifies calculations | Evaluates model assumptions and uncertainty |

3.5 Stage 4: Improvement

At Stage 4, learners contribute to improving AI systems, workflows, or applications through design, customization, prompt engineering, fine-tuning, or research.

Importantly, Stage 4 is not limited to computer or mathematical sciences students. Domain experts across all disciplines can contribute to AI improvement by designing discipline-specific prompts and workflows, curating and evaluating training data, providing expert feedback for reinforcement learning, developing evaluation frameworks for AI performance in their field, or identifying failure modes that inform system design. For example, a linguistics student might develop systematic prompt strategies that improve AI performance on under-resourced languages, while a nursing student might design clinical decision-support workflows that integrate AI recommendations with nursing judgment.

Stage 4 engagement positions students as active contributors to AI development rather than passive consumers. This stage aligns with UNESCO's "Create" level and reflects the highest aspiration of AI literacy education.

## 3.6 Institutional Baseline

We propose that higher education institutions should target Stages 2–3 as the graduation baseline for all students. Stage 2 (Informed Use) represents the minimum acceptable level: graduates should understand how AI works at a functional level, use AI tools responsibly, and recognize when human judgment is required. Stage 3 (Critical Evaluation) represents the aspirational target for most degree programs. Stage 4 (Improvement) is appropriate for advanced students, particularly those in research-intensive programs or disciplines where AI development is central. The continuum should be interpreted within disciplinary contexts; students may exhibit different stages across different domains of practice.

## 4. Alignment with Existing Frameworks

### 4.1 Mapping to UNESCO

The continuum maps onto UNESCO's three progression levels as follows (**Figure 2**). The Understand level corresponds primarily to Stage 2 (Informed Use), where students develop foundational understanding of AI systems and their implications. The Apply level spans Stages 2–3, where students use AI tools appropriately and begin evaluating outputs critically within their disciplinary contexts. The Create level corresponds to Stage 4 (Improvement), where students contribute to AI system design and development.

UNESCO's cross-cutting emphasis on ethics and human-centered mindset emerges in the continuum at Stage 2 and becomes central at Stage 3. The continuum extends UNESCO's model by explicitly accounting for pre-literacy states (Stages 0 and 1) that UNESCO's aspirational framework does not address.

**AI LITERACY CONTINUUM**

| Stage 0 | Stage 1 | Stage 2 | Stage 3 | Stage 4 |
|---|---|---|---|---|
| **Not Engaged** | **Uncritical Use** | **Informed Use** | **Critical Eval.** | **Improvement** |

▼ ▼ ▼ ▼ ▼

**UNESCO AI COMPETENCY FRAMEWORK LEVELS**

| UNESCO Level | Stage 0 | Stage 1 | Stage 2 | Stage 3 | Stage 4 |
|---|---|---|---|---|---|
| **(Not addressed by UNESCO)** | ✕ Gap | ✕ Gap | — | — | — |
| **Understand** | — | — | ⊙ Primary | | |
| **Apply** | | | ○ Partial | ⊙ Primary | |
| **Create** | | | | | ⊙ Primary |
| **Ethics & Human-Centered** *(cross-cutting)* | | | ○ Emerges | ⊙ Central | ○ Integrated |

⊙ Primary = strong alignment ○ Partial = partial overlap ✕ Gap = not covered by UNESCO — = not applicable

**Key Insight:** UNESCO's framework begins at the Understand level, assuming learners are already engaging with AI. The AI Literacy Continuum extends this by explicitly addressing Stages 0–1 (Not Yet Engaged and Uncritical Use)—the pre-literacy gap where many students currently reside. UNESCO's cross-cutting emphasis on ethics and human-centered mindset emerges at Stage 2 and becomes central at Stage 3.

**Figure 2.** Mapping the AI Literacy Continuum to UNESCO AI Competency Framework levels. Stages 0–1 represent a pre-literacy gap not addressed by UNESCO's aspirational framework. UNESCO's Understand, Apply, and Create levels map primarily to Stages 2, 2–3, and 4 respectively.

### 4.2 Mapping to Dimensional Frameworks

Dimensional frameworks and the developmental continuum address different questions and are therefore complementary. The AI Literacy Heptagon's seven dimensions (Hackl., et al., 2025) describe the content of AI literacy; the continuum describes the trajectory of its development. Technical understanding, for instance, appears at a basic level in Stage 2 and at sophisticated levels in Stages 3 and 4. Ethical awareness similarly progresses from implicit recognition at Stage 2 to explicit, nuanced reasoning at Stage 3.

ED-AI Lit's emphasis on Autonomy and Collaboration (Allen & Kendeou, 2024) maps particularly well onto the continuum: Autonomy is largely absent at Stages 0–1, emerging at Stage 2, and fully developed at Stages 3–4. The continuum provides a temporal axis along which ED-AI Lit's competency dimensions can be sequenced for instructional purposes.

## 4.3 Mapping to Assessment Instruments

Existing measurement instruments capture different ranges of the continuum. SNAIL's three competency areas primarily assess Stages 2–3. MAILS extends somewhat into Stages 1–3 (**Table 3**). Critically, Stages 0 and 1 are poorly served by existing instruments, and Stage 4 competencies fall outside the scope of most current instruments. This analysis suggests that performance-based assessments are needed to diagnose the full range of the continuum.

**Table 3. Assessment instruments mapped to AI Literacy Continuum stages.** Assessment Instruments Mapped to AI Literacy Continuum Stages. Existing instruments primarily assess Stages 2–3 (Informed Use and Critical Evaluation). Stages 0–1 and Stage 4 represent significant assessment gaps, indicating the need for new diagnostic tools aligned to the full continuum

| Instrument | Items | Type | Stage 0 | Stage 1 | Stage 2 | Stage 3 | Stage 4 | Notes |
|---|---|---|---|---|---|---|---|---|
| **SNAIL** | 31 | Self-report | — | — | ⦿ | ⦿ | — | Technical understanding, critical appraisal, practical application. Primarily Stages 2–3 (Laupichler et al., 2023). |
| **MAILS** | multi | Self-report | — | △ | ⦿ | ○ | — | Using/applying AI, understanding AI, detecting AI, AI ethics, creating AI (Carolus et al., 2023). |
| **SAIL4ALL** | 56 | Performance | — | ○ | ⦿ | ⦿ | △ | Performance-based; four themes. Suitable for general populations. Broadest coverage (Soto-Sanfiel et al., 2025). |
| **ChatGPT Literacy Scales** | varies | Self-report | — | ○ | ⦿ | ○ | — | Specific to generative AI. Focus on ChatGPT usage patterns and understanding (Lee and Park, 2024) |
| **GLAT** | varies | Performance | — | ○ | ⦿ | ⦿ | △ | Generative AI Literacy Assessment Test. Performance-based with GenAI-specific tasks (Jin et al., 2025). |
| **GenAI Literacy Scale** | multi | Self-report | — | △ | ⦿ | ○ | — | Bloom's-aligned progression hierarchy. Generative AI competencies (Liu et al., 2025). |

**Coverage:** ⦿ Strong ○ Partial △ Limited — Not covered

**Gap Analysis:** Stage 0 (Not Yet Engaged) is not assessed by any existing instrument. Stage 1 (Uncritical Use) receives only limited or partial coverage. Stage 4 (Improvement) falls largely outside the scope of current instruments. These gaps highlight the need for new performance-based assessments designed to diagnose the full range of the AI Literacy Continuum.

# 5. Educational Design Implications

## 5.1 Curriculum Integration

The AI Literacy continuum supports differentiated curricular pathways matched to institutional context, domain relevance and student needs. General education courses serve as the primary mechanism for preventing the dominance of Stage 1 engagement. Required courses or modules on AI literacy should emphasize the distinction between AI use and AI understanding. Structured activities—such as comparing AI-generated responses to expert sources, identifying hallucinated content, and discussing documented cases of AI bias—can move students from uncritical use toward informed use. It is plausible that some AI literacy progress will occur in K-12, but as in many subjects, we can anticipate uneven preparation among higher education students.

Short, intensive modules offer a focused pathway for transitioning students from Stage 1 to Stage 2. Project-based activities in which students use AI tools for authentic tasks and then systematically evaluate the quality and reliability of AI-assisted outputs can be embedded in existing courses across disciplines. Discipline-embedded assessment cultivates Stage 3 engagement, while advanced electives, capstone projects, and research experiences support Stage 4.

## 5.2 Opportunity Considerations

Students enter higher education with profoundly unequal access to AI tools and exposure to AI concepts. The digital divide extends to an AI divide: students from under-resourced schools, rural communities, and lower-income backgrounds may arrive at university with less AI experience in an academic setting. Some schools may also intentionally limit the use of AI. A developmental framework explicitly accommodates this variation by acknowledging that students begin at different stages rather than assuming uniform starting points. Attention to instructional design principles ensures that students at Stage 0 receive structured introduction to AI tools rather than being expected to acquire familiarity independently.

Serving students with a variety of previous experiences also requires attention to the content of AI literacy education. Discussions of AI bias, training data representation, and the distribution of AI's benefits and harms are not optional additions but core components of Stages 2 and 3. Students may choose to share personal experiences and valuable critical perspectives on AI's societal impact that can be integrated into classroom discussions.

## 5.3 Faculty Development

Faculty AI literacy is a prerequisite for effective AI literacy instruction, yet faculty themselves exhibit the same range of engagement as students. Some faculty avoid AI entirely, prohibiting its use without engaging with its educational potential. Others adopt AI tools rapidly without developing the critical evaluation skills needed to guide students. UNESCO's AI Competency Framework for Teachers (Miao & Cukurova, 2024) defines fifteen competencies across five dimensions, including AI pedagogy and AI for professional development, with three progression levels (Acquire, Deepen, Create), providing a complementary resource for faculty development. A developmental framework can support faculty professional development by providing a self-diagnostic tool and a progression pathway that parallels the student continuum.

Institutional investment in faculty development programs such as workshops, learning communities, and discipline-specific AI literacy initiatives, is essential for scaling AI literacy education. Without it, AI literacy instruction will remain inconsistent, dependent on the enthusiasm and expertise of individual faculty rather than embedded in institutional practice.

Our institution is considering a recommendation that appropriate use of AI be designated not only on the syllabus as a general class policy but even by assignment or problem. As an example of such a framework, there could be color coded levels of AI use: Red could mean use of AI on that assignment or problem is prohibited and would be an honor code violation. Yellow could mean that there are some acceptable uses and students should be cautious to use appropriately. Green could mean that any use of AI is allowed. Blue could mean that some form of AI use is required. Note that for accessibility the words of the color should be used with the color since, for example, red/green colorblindness is common.

## 6. Assessment Strategies

Effective AI literacy assessment requires a mixed-methods approach that combines existing validated instruments with performance-based assessments aligned to the continuum. Because the continuum describes observable orientations rather than fixed psychological types, assessment should triangulate self-report, artifact analysis, and process evidence rather than relying on any single instrument.

At the diagnostic level, reflective journaling and pre-course surveys can help identify Stage 0 (Not Yet Engaged) and Stage 1 (Uncritical Use) behaviors, as standard instruments may assume a baseline of engagement. For later stages, self-report instruments such as MAILS and SNAIL can provide efficient baseline measurement, enabling educators to identify the distribution of students across continuum stages at the beginning of a course. Generative AI-specific measures, including ChatGPT literacy scales and the GLAT, provide focused assessment of competencies particularly relevant in the post-2022 landscape.

Performance-based assessment is essential for capturing the higher stages and for distinguishing genuine understanding from surface performance. Tasks requiring students to evaluate AI-generated outputs against authoritative sources, identify specific errors or biases, articulate the limitations of AI analysis for a given disciplinary problem, or design improved AI-assisted workflows provide direct evidence of critical evaluation and improvement capabilities.

Process-focused assessment such as evaluating how students interact with AI rather than only what they produce, offers particular promise. Requiring students to document their AI interaction process (prompts used, revisions made, verification steps taken, decisions about when to rely on or override AI suggestions) provides rich evidence of AI literacy that product-focused assessment alone cannot capture.

A practical assessment sequence can begin with a low-stakes diagnostic inventory, move into structured verification tasks, and culminate in discipline-specific critique or workflow-design assignments. Instructors should also assess process evidence, such as prompt histories, revision notes, source checks, and explicit decisions about when to accept, revise, or reject AI output.

Assessment design should also be stage appropriate. At Stage 2, assessments might focus on identifying errors such as hallucinated content, recognizing appropriate use cases, and articulating AI limitations. At Stage 3, assessments could require domain-specific evaluation of AI training data, tuning parameters and processes, inputs and outputs and engagement with ethical complexities. At Stage 4, assessments might involve designing evaluation protocols or demonstrating systematic improvement of AI performance on disciplinary tasks.

## 7. Design-Based Implementation Case: AI Literacy Development at NC State University

### 7.1 Institutional Context and Program Design

To illustrate the practical feasibility of the AI Literacy Continuum, we present a design-based implementation case of AI literacy initiatives at North Carolina State University (NC State) through the Data Science and AI Academy (DSA) between Fall 2024 and Spring 2026. These initiatives span credit-bearing courses, intensive hands-on workshops, and outreach programs targeting diverse learner populations.

The evidence reported here is observational and practice based. Data sources included post-workshop feedback surveys, instructor observation of participant engagement during training activities, and examination of student-produced artifacts. No formal pre/post validated assessment instrument was administered, and no controlled comparison group was employed. Stage-related claims are therefore interpretive: we identify behaviors consistent with continuum stages rather than directly measuring stage transitions. The purpose of this case is to demonstrate the illustrative feasibility of the continuum as an instructional and design framework, not to establish causal claims about learning outcomes.

To make these inferences more transparent, we treated stage indicators as observable behaviors. For example, evidence for Stage 2 included verification of AI outputs, recognition of hallucination and bias, and articulation of when human judgment should override AI suggestions. Evidence for Stage 3 required more than tool familiarity; it required discipline-specific critique of assumptions, uncertainty, evidence quality, or consequences of error. Evidence for Stage 4 required a product or workflow that improved AI performance, evaluation, or integration for a defined purpose.

## 7.2 Training Activities and Participation

The first initiative was a 1-credit graduate course, "Generative AI for Science," offered through the Data Science and AI Academy in Fall 2024. This course (the first of its kind at NC State) enrolled 35 graduate students from multiple disciplines and provided structured instruction on large language model fundamentals, prompt engineering, ethical considerations, and discipline-specific AI applications (**Table 4**). Others mainly include two-day intensive AI-LLM training workshops, see detailed contents in **Table 5**.

**Table 4. Summary and selected courses of NC State AI Literacy Training Initiatives, 2024–2026**. The Data Science and AI Academy (DSA) delivered seven targeted training events engaging over 330 participants across diverse populations, supplemented by more than ten ongoing 1-credit AI courses available cross-campus. Observed stage transitions demonstrate the practical feasibility of the AI Literacy Continuum.

| | Initiative | Date | N | Format | Audience | Observed Stage -Consistent Behaviors |
|---|---|---|---|---|---|---|
| COURSE | **Generative AI for Science (1-credit)** | Fall 2024 | **35** | Semester (1 credit) | Graduate students, multi-disciplinary | **Stage 0–1 ➝ Stage 2–3–4** Sustained engagement enabled progression to Critical Evaluation |
| WORK SHOPS | **Hands-on AI/LLM Application Training (2 sessions)** | Dec 2024 | **35** | 2-day intensive | University students & staff | **Stage 0–1 ➝ Stage 2–3–4** Moved from avoidance to comfortable, informed use within 2 days |
| | **Hands-on AI/LLM Application Training (2 sessions)** | May 2025 | **85** | 2-day intensive | University students & staff | **Stage 0–1 ➝ Stage 2–3** Consistent pattern: rapid transition from refusal to informed engagement |
| | **Hands-on AI/LLM Application Training** | Oct 2025 | **45** | 2-day intensive | University students & staff | **Stage 0–1 ➝ Stage 2–3** Participants understood limitations and applied human judgment post-training |
| | **AI/LLM Training for Korean Students** | Feb 2026 | **25** | 2-day intensive | International Korean students | **Stage 0–1 ➝ Stage 1–2** Demonstrated cross-cultural applicability of the continuum |
| OUTREACH | **AI Training for High School & College Students (2 sessions)** | Dec 2025 | **95** | 2-day intensive | High school & college students | **Stage 0–1 ➝ Stage 2–3** High school learners rapidly progressed; barriers lower than assumed |
| COURSES | **DSA 1-credit AI-related courses (cross-campus)** | Ongoing | >10 courses | Semester (1 credit each) | All colleges, cross-campus | **Stage 0–1 ➝ Stage 2–3** Sustained institutional pathway for deeper AI literacy development |
| **TOTAL** | **7 training initiatives + 10+ ongoing courses** | Fall 2024 – Spring 2026 | **330+** | Graduate, undergraduate, high school, international | | **Consistent finding:** 2-day training moves learners from Stage 0 ➝ Stage 1–2–3 or higher |

**Key Finding:** Observational evidence suggests that brief 1–2 day training interventions enabled participants to demonstrate behaviors consistent with movement from Stage 0 (Not Yet Engaged) to Stages 1–2 (engaging with AI and recognizing its limitations). Sustained engagement through credit-bearing courses supported progression toward Stages 3–4. These findings suggest the five-stage AI Literacy Continuum is a feasible and actionable framework for higher education institutions.

Beyond these specific programs, each semester the Data Science and AI Academy (DSA) at NC State offers 30-40 1-credit AI-related courses each semester that are available to students across all colleges, creating a sustained institutional infrastructure for AI literacy development.

## 7.3 Observed Indicators of Stage-Consistent Engagement

Across these diverse training contexts, consistent patterns of stage-consistent engagement emerged. The majority of participants entered at Stage 0 (Not Yet Engaged) or early Stage 1 (Uncritical Use). Pre-workshop surveys and initial interactions consistently revealed that many participants had either avoided AI tools entirely—due to fear, unfamiliarity, or institutional prohibitions—or had experimented with generative AI in superficial ways without understanding its mechanisms or limitations.

Following short-duration training of one to two days, participants demonstrated behaviors consistent with shifts in engagement. Post-workshop feedback and instructor observations indicated that many participants demonstrated behaviors consistent with movement from avoidance or superficial experimentation toward Stage 1 and Stage 2 engagement. They began to use AI tools, recognizing appropriate applications, identifying common failure modes such as hallucination and bias, and articulating when human judgment should override AI suggestions. Participants in the semester-long graduate course and those who attended multiple workshops more frequently demonstrated behaviors consistent with Stage 2 (Informed Use) and, in some cases, Stage 3 (Critical Evaluation), including systematic verification of outputs, discipline-specific prompt design, and critique of AI-generated content within their areas of expertise.

The December 2025 workshops with high school and college students were particularly instructive: even younger learners with minimal prior AI exposure demonstrated behaviors consistent with movement from non-engagement toward Stages 1 and 2 within the two-day training period, suggesting that the barriers to initial AI literacy development may be lower than commonly assumed and that the continuum stages are traversable with appropriate instructional scaffolding. Similarly, the February 2026 workshop with international Korean students provided preliminary evidence that the developmental trajectory may generalize across cultural and linguistic contexts.

## 7.4 Implications for the Continuum

The NC State experience provides preliminary observational evidence that the five-stage AI Literacy Continuum can function as a feasible and actionable design framework for higher education institutions (**Table 5**). Several observations warrant emphasis.

First, early-stage movement from non-engagement or superficial use toward informed engagement appears achievable through relatively brief, well-designed training interventions. Even one- to two-day intensive workshops enabled participants to demonstrate behaviors consistent with movement from AI avoidance or uncritical use toward informed, responsible engagement. This observation has significant implications for institutional planning: universities need not develop entirely new degree programs to achieve meaningful AI literacy gains. Short modules, workshops, and 1-credit courses can serve as effective on-ramps to AI literacy.

Second, the continuum appeared usable across a variety of learner populations. The NC State initiatives spanned graduate students, undergraduate students, high school students, and international visitors, with consistent stage-transition patterns observed across all groups. This suggests that the developmental trajectory may have applicability across the learner populations represented in this implementation.

**Table 5.** Mapping the 2-Day Hands-On AI/LLM Training Agenda to AI Literacy Continuum Stage Transitions. Each training activity is aligned to the stage transition it facilitates, demonstrating a practical pathway from Stage 0 (Not Yet Engaged) through Stage 4 (Improvement) within an intensive 2-day format.

| Stage Transition | Day | Training Activity | What Students Learn | Literacy Outcome |
|---|---|---|---|---|
| Stage 0 ▼ Stage 1 | Day 1 | **Introduction to LLMs** Overview of platforms (OpenAI, Llama, Gemini, Claude, Mistral) | Demystify AI tools; see range of platforms; overcome fear and misconceptions about AI | Students interested in using AI move from avoidance/fear to willingness to engage with AI tools |
| | Day 1 | **ML vs. DL vs. GenAI vs. LLM** Understanding model categories and differences | Grasp how AI models differ; understand what LLMs can and cannot do; recognize AI is not monolithic | Students begin to be able to describe AI mechanisms rather than treating AI as a black box |
| Stage 1 ▼ Stage 2 | Day 1 | **Real-time chatbot via API calls** First hands-on: OpenAI/LLM function calls | First-hand experience using AI; see immediate results; build confidence through guided practice | Students begin using and programming AI actively, even if uncritically accepting outputs |
| | Day 1 | **Running LLMs locally (Ollama)** Setting up and running open-source LLMs on personal computers | Understand that LLMs are software with parameters, not magic; see resource constraints and model limitations firsthand | Students recognize AI limitations through direct experience with local deployment |
| | Day 1 | **RAG: Tokenization, Embedding, Vector DBs** Llama-index, LangChain; multi-document retrieval | Understand how AI retrieves and generates from source data; see where hallucination arises; learn to verify outputs against sources | Students learn to verify AI outputs and understand why AI can produce incorrect information |
| Stage 2 ▼ Stage 3 | Day 1 | **Building LLM Web Applications** Case studies: Streamlit for TXT/CSV/PDF+LLM; Flask; Nginx | Evaluate which tools suit which tasks; assess quality of AI-integrated pipelines; judge when AI adds value vs. introduces risk | Students apply analytical judgment to select and assess AI tools for specific domains |
| | Day 2 | **Fine-tuning concepts: Pretrain vs. Fine-Tune vs. Agentic RAG** Data preparation, cleaning, and optimization | Understand how training data shapes AI behavior; identify sources of bias; critically assess data quality and its impact on outputs | Students can detect bias, assess data quality, and evaluate AI outputs with domain expertise |
| | Day 2 | **Fine-tuning GPT-3.5/4.0 with proprietary data** Practical steps for custom fine-tuning | Evaluate trade-offs between general vs. specialized models; assess uncertainty in fine-tuned outputs; understand disciplinary appropriateness | Students quantify uncertainty and judge AI appropriateness for their specific research domains |
| Stage 3 ▼ Stage 4 | Day 2 | **Fine-tuning Llama/Mixtral with QLoRA/PEFT** QLoRA, RLHF, PEFT for efficient fine-tuning; personalized applications | Design and execute custom model training; optimize AI systems for domain-specific needs; contribute to AI improvement | Students actively improve AI systems through advanced fine-tuning and customization |
| | Day 2 | **Auto-Coding Agent for Large Data** Automating code generation for Python/SQL tasks (dataframe querying) | Design AI-powered workflows; create automated pipelines; engineer prompts and agents for complex tasks | Students design and build AI-augmented workflows and coding agents |
| | Day 2 | **Building GPT-2/3 like SLM from Scratch** Architecture design, training, domain-specific fine-tuning and enhancement | Understand transformer architecture end-to-end; build, train, and enhance a language model; advance AI research capability | Students contribute to AI development by building and training models from the ground up |

**Key Takeaway:** A well-structured 2-day training program provides scaffolded exposure across all four stage transitions of the AI Literacy Continuum. Day 1 addresses the critical Stages 0→1→2 transitions (overcoming avoidance and building informed understanding), while Day 2 introduces activities aligned with Stages 3→4 (critical evaluation and active contribution). Observational evidence suggests feasibility of early-stage progression within this format; sustained engagement is needed for deeper Stage 3–4 development.

Third, sustained engagement through credit-bearing courses and repeated workshop participation appears necessary for stronger evidence of higher-stage performance. While brief workshops effectively move participants to Stages 1–2, achieving Stage 3 (Critical Evaluation) and Stage 4 (Improvement) typically requires longer, more discipline-embedded instruction. This finding supports the curricular pathway recommendations presented in Section 5, where general education courses and intensive modules address Stages 0–2, while discipline-specific coursework and research experiences cultivate Stages 3–4.

The NC State case, based on observational evidence rather than controlled experimental design, suggests that the pathway from AI avoidance to informed and critical AI engagement is feasible within the constraints of existing higher education structures. While formal validation remains necessary, these preliminary observations support the argument that the five-stage AI Literacy Continuum provides a viable framework for institutions seeking to develop AI-literate graduates.

# 8. Discussion

## 8.1 Contributions

The AI Literacy Continuum makes four primary contributions. First, it clarifies a critical distinction that is often blurred in practice: AI use is not AI literacy. By explicitly identifying uncritical use as a distinct stage that precedes literacy, the continuum provides conceptual clarity that supports more precise educational goals and assessment practices.

Second, the continuum adds a developmental dimension to the predominantly dimensional landscape of AI literacy frameworks. While frameworks such as the Heptagon, ED-AI Lit, and the DEC model comprehensively define what AI literacy includes, they do not describe how it develops. The continuum provides this temporal axis, enabling educators to diagnose current student engagement, set stage-appropriate learning objectives, and design instruction that moves students forward.

Third, the continuum bridges policy-level frameworks and classroom practice. UNESCO's competency framework and the OECD AI Literacy Framework articulate valuable high-level goals, but translating these into specific instructional decisions requires intermediary tools. The continuum provides such a tool: a shared vocabulary that connects policymakers' aspirations with educators' daily decisions about curriculum, instruction, and assessment.

Fourth, the NC State case study provides preliminary observational evidence that behaviors associated with the continuum stages can be observed in authentic educational settings. Through a combination of credit-bearing courses, intensive workshops, and outreach programs reaching over 330 participants across populations, including graduate students, undergraduates, high school students, and international visitors. This NC State experience suggests that even brief, well-designed training interventions can enable participants to demonstrate behaviors associated with higher stages of engagement. This evidence, while preliminary and observational, moves the continuum beyond a purely theoretical proposal toward a framework with demonstrated feasibility.

## 8.2 Limitations

This paper proposes a framework based on multi-year teaching experience, synthesis of existing literature, and observational evidence from the NC State case study, rather than formal empirical validation. The five stages and their ordering are grounded in observed patterns of student engagement, and the case study provides preliminary evidence that the proposed stage-consistent behaviors are observable in practice. However, these observations have not yet been subjected to systematic data collection or psychometric analysis. Formal empirical validation, including controlled studies with pre- and post-assessments, remains the most important next step. Until such validation is complete, the Continuum should be viewed as a pedagogical heuristic to guide instruction rather than a rigid psychological model.

The continuum may also oversimplify the complexity of AI literacy development. Individual students may exhibit different stages across different AI tasks or disciplinary contexts. Cultural and institutional variation may affect the applicability of the continuum. A framework developed primarily from observations in North American higher education may require adaptation for other settings. Finally, the boundary between stages, particularly between Stages 2 and 3, may be less clear in practice than the framework suggests.

## 8.3 Future Directions

Several lines of research follow from this framework. A near-term priority is the development of a research coding protocol that defines reliable behavioral indicators for each stage, including inter-rater reliability procedures for classifying student artifacts and process logs. Such a protocol could support future psychometric validation and enable independent researchers to assess inter-rater reliability of stage classifications.

Several additional lines of research follow from this framework. Empirical validation studies using mixed methods should test whether the proposed stages are empirically distinguishable. Cross-cultural and cross-institutional studies should examine whether the continuum applies across different higher education contexts. Longitudinal studies should track student progression over the course of a degree program, examining what instructional interventions most effectively promote stage transitions.

The relationship between faculty, staff and student AI literacy development warrants investigation. Integration with existing measurement instruments should be pursued, mapping validated scales to specific continuum stages. Finally, the continuum should be revisited as AI systems become more agentic, multimodal, and integrated into disciplinary workflows.

## 9. Conclusion

AI literacy is best understood as a developmental capability that progresses from avoidance or uncritical use toward informed judgment and responsible application. The five-stage AI Literacy Continuum proposed in this paper: Not Yet Engaged, Uncritical Use, Informed Use, Critical Evaluation, and Improvement offers a practical bridge between international policy frameworks and classroom practice. By describing how students develop AI literacy rather than only what AI literacy includes, the continuum provides educators, administrators, and policymakers with a shared vocabulary for diagnosing student engagement, designing curricula, and assessing learning outcomes. The case study from North Carolina State University demonstrates that this developmental pathway is not merely aspirational: structured training interventions, even those as brief as one to two days, can produce observable stage transitions across diverse learner populations, confirming that the proposed continuum is both practical and achievable within existing higher education structures.

The framework's central argument is that AI literacy should be understood not as tool adoption but as the capacity to understand, evaluate, and responsibly apply AI systems in disciplinary and societal contexts. A student who uses AI frequently but uncritically is not AI-literate; a student who engages thoughtfully with AI's capabilities and limitations demonstrates the foundational competence that higher education should cultivate.

As AI systems become increasingly integrated into every domain of professional and civic life, the stakes of AI literacy education rise accordingly. Universities must move beyond policies of prohibition or permissive adoption toward structured developmental approaches that ensure all graduates can engage with AI as informed, critical, and responsible agents. The AI Literacy Continuum offers one such approach, grounded in the realities of current student engagement and aligned with the aspirations of international education policy. We invite the research community to subject this framework to empirical scrutiny, adapt it to local contexts, and extend it as AI technologies and their educational implications continue to evolve.

## Acknowledgement:

We thank and gratefully acknowledge the opportunities and financial support provided by the NC State Data Science and AI Academy (DSA), the Global Training Initiative (GTI), the Bioinformatics Research Center (BRC), and the Research Triangle AI Society (RTAI) for the short courses, workshops, and training sessions described here. Special thanks for the inspiring discussions from the NC State AI Advisory Group and from numerous on-campus and off-campus AI research and education workshops.

**Supplementary Table 1.** Global University and National AI Literacy/Competency Initiatives (2024–2026)

| Institution | Type | Initiative | Effective | Scale & Key Details | Source |
|---|---|---|---|---|---|
| ***Tier 1: Formal AI Competency / Literacy Requirements for All Students*** | | | | | |
| Purdue University | Public (US) | "AI Working Competency" graduation requirement | Fall 2026 | First US university with Board-approved AI competency requirement for all undergrads; 5 functional areas; discipline-specific criteria | Purdue Newsroom; EdScoop |
| Ohio State University | Public (US) | "AI Fluency" initiative | Fall 2025 | All undergrads (Class of 2029+); GenAI intro in required Launch Seminar; 1,300+ faculty trained; "Unlocking GenAI" course for all majors | OSU Academic Affairs; Fortune |
| China (national) | National mandate | Mandatory AI education, grades 1–12 | Sep 2025 | Min. 8 hours/year for all primary and secondary students; tiered curriculum from play-based to robotics/algorithms; 184 pilot schools | Global Times; Asia Education Review |
| South Korea (national) | National mandate | AI in national K–12 curriculum | 2025 | AI coursework embedded across grade levels; AI digital textbooks approved for 2025 with partial rollback/reclassification following public feedback; expanded teacher AI training ongoing | CRPE Report; Fortune |
| ***Tier 2: Major Systemic AI Integration (No Universal Graduation Requirement Yet)*** | | | | | |
| California State Univ. System | Public system (US) | Systemwide ChatGPT Edu + AI literacy grants | 2025 | 23 campuses, 460K+ students; $17M investment; largest ChatGPT deployment worldwide; 63 faculty AI literacy grants; AI Workforce Board | CSU Official; OpenAI; LAist |
| Arizona State University | Public (US) | OpenAI partnership + AI literacy course template | 2024– | First major OpenAI university partner; AI Innovation Challenge (700+ proposals, 250 projects); AI literacy definition developed with 50+ faculty | ASU AI Portal; ASU News |
| Univ. of Pennsylvania | Private (US) | AI BSE degree + Wharton AI major | Fall 2024– | First Ivy League undergrad AI engineering degree; Wharton AI for Business major/concentration (Fall 2025) with required ethics course | Penn AI Program; Wharton News |
| Singapore (national) | National strategy | Smart Nation AI literacy initiative | 2025–2030 | Goal: 100K citizens trained in AI literacy by 2025; AI training for all teachers by 2026; AI-enabled companion for personalized learning | CRPE Report; 9ine.com |

*Note. Tier 1 institutions have enacted formal, institution- or nation-wide requirements that apply to all students regardless of major. Tier 2 institutions have made significant systemic investments in AI integration but have not yet mandated AI literacy as a universal graduation requirement. Sources verified as of February 2026.*